\documentclass[letterpaper,twocolumn,10pt]{article}
\usepackage{zhanggroup}

\usepackage{xspace}
\usepackage{amsmath}
\usepackage{graphicx}
\usepackage{caption}
\usepackage{subcaption}
\usepackage{float}
\newcommand{\mypara}[1]{\noindent{\bf {#1}.}\xspace}
\newcommand{\geoagent}{GEO-Detective\xspace}
\begin{document}

\date{}

\title{\bf GEO-Detective: Unveiling Location Privacy Risks in Images with LLM Agents}

\author{
Xinyu Zhang\textsuperscript{1}\ \ \
Yixin Wu\textsuperscript{1}\ \ \
Boyang Zhang\textsuperscript{1}\ \ \
\\
Chenhao Lin\textsuperscript{2}\ \ \
Chao Shen\textsuperscript{2}\ \ \
Michael Backes\textsuperscript{1}\ \ \
Yang Zhang\textsuperscript{1}
\\
\\
\textsuperscript{1}\textit{CISPA Helmholtz Center for Information Security} \ \ \ 
\textsuperscript{2}\textit{Xi'an Jiaotong University} \ \ \
}

\maketitle

\begin{abstract}
Images shared on social media often expose geographic cues. 
While early geolocation methods required expert effort and lacked generalization, the rise of Large Vision Language Models (LVLMs) now enables accurate geolocation even for ordinary users. 
However, existing approaches are not optimized for this task. 
To explore the full potential and associated privacy risks, we present Geo-Detective, an agent that mimics human reasoning and tool use for image geolocation inference. 
It follows a procedure with four steps that adaptively selects strategies based on image difficulty and is equipped with specialized tools such as visual reverse search, which emulates how humans gather external geographic clues. 
Experimental results show that GEO-Detective outperforms baseline large vision language models (LVLMs) overall, particularly on images lacking visible geographic features. 
In country level geolocation tasks, it achieves an improvement of over 11.1\% compared to baseline LLMs, and even at finer grained levels, it still provides around a 5.2\% performance gain. 
Meanwhile, when equipped with external clues, GEO-Detective becomes more likely to produce accurate predictions, reducing the “unknown” prediction rate by more than 50.6\%. 
We further explore multiple defense strategies and find that Geo-Detective exhibits stronger robustness, highlighting the need for more effective privacy safeguards. 
\end{abstract}

\section{Introduction}
\label{sec:intro}

Users frequently share images on social media that contain geographic cues such as landmarks, architectural styles, and signage. 
These cues have long posed privacy risks, enabling doxing attacks that infer and expose private geolocation~\cite{RB19,OSF17}. 
Importantly, many users are unaware that such seemingly innocuous visual details can leak sensitive location information. 
Even users who deliberately remove metadata (e.g., EXIF) to protect their privacy may still be identified when determined adversaries analyze the visual content itself. 
Similar incidents~\cite{bbc_hidden_fingerprint,guardian_mcafee_location} have also been frequently reported in the news, underscoring the subtle yet persistent nature of these privacy risks. 
However, such attacks traditionally required specialized expertise and manual effort, which constrained their scale. 
Recent advances in large vision language models (LVLMs)~\cite{JYXCPPLSLD21,RKHRGASAMCKS21,LLWL23} have introduced reasoning capabilities over visual content. 
When combined with tools like web search, this capability greatly lowers the barrier to geolocation inference. 
Ordinary users can now infer image geolocations with minimal effort, achieving accuracy that once demanded expertise. 
This shift has amplified privacy risks, enabling doxing at a greater scale and with higher precision, resulting in severe privacy threats such as identity exposure and behavioral tracking. 

Early geolocation methods framed the task as image classification, partitioning the globe into grid cells and predicting the most likely cell for a given image~\cite{WKP16, MPE18}. 
While effective for coarse localization, these models lacked fine grained resolution and interpretability, properties essential for privacy attacks in the real world. 
More recent work has explored leveraging large vision language models (LVLMs) to generate descriptive reasoning. 
Some systems even adopt tool augmentation~\cite{WYQWTD23, SSTLLZ23,YZYDSNC23}, enabling external tools such as web search or code execution to support inference. 
However, these approaches are not specifically designed for geolocation and thus fall short of simulating an adversary intent on maximizing inference success. 

To bridge the gap, we propose \geoagent, an autonomous agent that combines LVLM reasoning with external tools to examine its geolocation capabilities and assess the resulting privacy implications. 
Our agent is designed to approximate how humans reason about geolocation tasks. 
Given an image, a person typically identifies distinctive geographic cues, such as landmarks, architectural features, or signage, and adjusts their effort based on task difficulty: a unique landmark allows quick inference, whereas sparse cues require deeper investigation and external resources. 
Inspired by this process, our agent adopts a similar adaptive approach. 
It first estimates difficulty from visible cues, then selects an appropriate strategy that may involve one or more tools. 
These include LVLMs based analysis, experience augmented prompting, geographic feature segmentation, and visual reverse search, which together complement LVLM capabilities by extracting salient cues, leveraging prior knowledge, and retrieving external evidence. 

In our evaluation, \geoagent outperforms the advanced LVLM o3 from OpenAI~\cite{o3}, especially under more challenging conditions. 
On average, it achieves about 3.0 to 5.0\% improvements over o3 across different difficulty levels. 
The gain is even more pronounced on GPT-4o~\cite{GPT4o}, where the agent reaches up to 8.0\% higher accuracy on difficult and very difficult images. 
With external geographic clues, the agent is more likely to make a prediction, cutting the ``unknown'' rate by more than half. 
This also raises serious privacy concerns. To assess possible mitigation strategies, we evaluate four defense mechanisms. 
Among them, adding a visible watermark stating that geolocation is not allowed proves highly effective, as both LLMs and agents refrain from making predictions. 
However, for the other defenses, the proposed agent demonstrates greater robustness compared to baseline LLMs, maintaining its ability to infer locations even under countermeasures. 

\noindent Overall, our contributions are summarized as follows:
\begin{itemize}
    \item We introduce \geoagent, an agent that mimics human reasoning for image geolocation, following a pipeline of four stages: visual analysis, strategy selection, result synthesis, and iterative refinement. 
    \item We equip \geoagent with several specialized tools, such as visual reverse search, which emulates how humans gather external geographic clues. 
    \item We demonstrate that \geoagent significantly amplifies geolocation privacy risks, especially on difficult cases, and evaluate four defense strategies to mitigate potential misuse. 
\end{itemize}

\section{Related Works}
\label{sec:related_works}

\mypara{Geolocation Inference}
The geolocation task aims to predict the location of a photo from its visual content. 
Early work~\cite{WKP16, SWSH18, SCA18,HE08,VJH17} framed it as a classification problem. 
Recent approaches adopt contrastive learning, exemplified by GeoCLIP~\cite{CNS23}, which aligns images and locations in a joint embedding space. 
Similar frameworks leverage large scale image text pretraining such as CLIP~\cite{RKHRGASAMCKS21} and ALIGN~\cite{JYXCPPLSLD21}, adapting them to encode geographic semantics through coordinate or textual prompts~\cite{JLLZWDHWWY24,LDDLZSZGL24,MCHDXR24}. 
A Vision Transformer encoder extracts image features, and a location encoder maps coordinates into vectors~\cite{CNS23,MLHSE23,HAS23}. 
Trained on the MP16 dataset~\cite{JLLZWDHWWY24} (4.72M geotagged images), GeoCLIP learns associations between landmarks, environments, and cultural cues. 
At inference time, a query image is embedded and compared to preencoded locations; the top match is returned, and similarity scores indicate geographic relevance. 
The similarity score is formally defined as follows:

\begin{equation}
\label{equ:geo_clip}
s_{\text{GeoCLIP}}(I, T) = \frac{f_{\text{img}}(I) \cdot f_{\text{text}}(T)}{\|f_{\text{img}}(I)\| \, \|f_{\text{text}}(T)\|},
\end{equation}

where \(f_{\text{img}}(\cdot)\) and \(f_{\text{text}}(\cdot)\) denote the GeoCLIP image and text location encoders, respectively. 
However, doxing style geolocation often requires iterative reasoning~\cite{LZLLZXX25,YPGT25} rather than a single step retrieval. 
An adversary may need to extract salient cues, disambiguate similar candidates, cross check external sources, and provide an explanation. 
Pure similarity search lacks this multiple steps of reasoning, dynamic tool integration, and structured evidence synthesis, limiting its automation in open world scenarios. 

\mypara{LVLMs and Geolocation Privacy}
With the rapid improvement of reasoning abilities in large vision language models (LVLMs), researchers have explored their potential in geolocation tasks. 
Jia et al.\ proposed a distance aware ranking method for global localization~\cite{JPGZL25}. 
Jiang et al.\ demonstrated that combining chain of thought reasoning with tool use helps models parse complex environments~\cite{JCZYZ25}. 
GeoComp~\cite{SYHTZCFGC25} and ETHAN~\cite{LDDLZSZGL24} introduced large scale datasets and structured reasoning frameworks, offering more systematic ways to evaluate geolocation performance. 
Recent works such as LLMGeo~\cite{WXKLFZ24} and GeoLocator~\cite{YWLSW24.0} benchmarked multimodal LVLMs on geolocation tasks in the wild, revealing that advanced reasoning capabilities can approach or even surpass human performance. 
These advances show that LVLMs achieve rapidly improving accuracy in geolocation. 
However, they also raise privacy concerns: multimodal LVLMs can infer a user's home address or even coordinates at the GPS level from everyday photos~\cite{LZLLZXX25}, and interactions with multiple turns often reveal more location information than single turn queries~\cite{MCHDXR24}. 
Prior studies on visual privacy leakage~\cite{HHLLWZ25,MLBFWW22,LDDLZSZG25.0,SKU24,MCHDXR24} and multimodal reasoning leakage~\cite{LZLLZXX25,CLZCW25,LZLYQ24,TVSV24} further emphasize that fine grained scene understanding may unintentionally expose sensitive identity or location data. 
These findings suggest that while stronger reasoning improves accuracy, it also increases the risk of exposing sensitive location data. 
Most of these works study privacy leakage from the model side, focusing on how multimodal models expose sensitive information during reasoning. 
In contrast, our work uses an agent based framework that simulates human interactions with large vision language models (LVLMs) and integrates external tools such as image retrieval and image segmentation to reproduce realistic geolocation reasoning. 
This approach allows us to study privacy leakage not only from the model itself but also from the way it is used, revealing new risks that arise when tools are combined with reasoning in real world geolocation tasks. 

\begin{figure*}[!t]
\centering
\includegraphics[width=2\columnwidth]{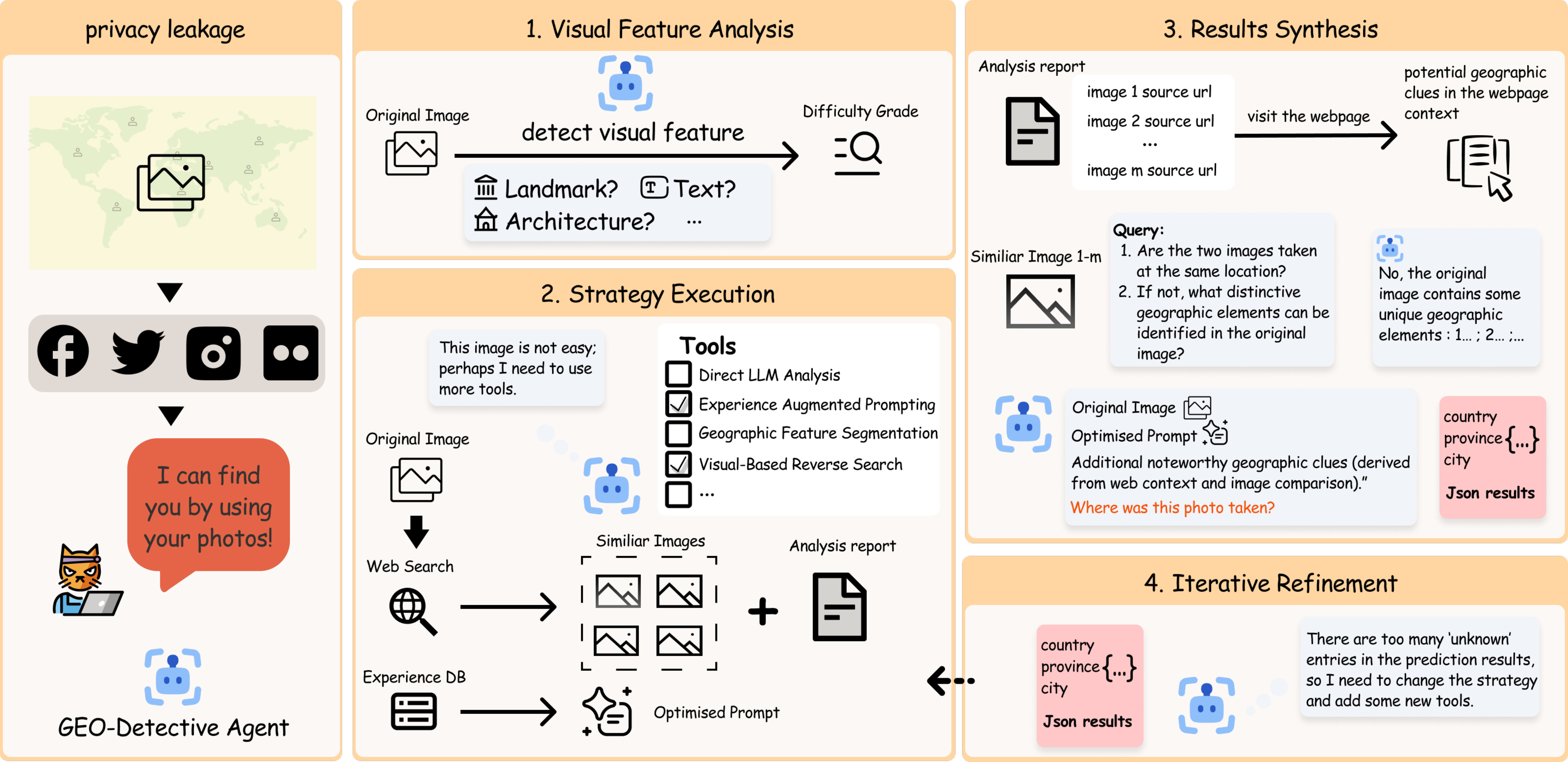}
\caption{System design of the \geoagent.
The agent takes images shared by users on online platforms as input and processes them through four stages, including visual feature analysis, strategy execution, result synthesis, and iterative refinement.
During this process, users' location information may be inferred, leading to potential privacy risks.}
\label{fig:architecture}
\end{figure*}

\section{Methodology}
\label{sec:methodology}

\subsection{Overview}
\label{sec:overview}

To support geolocation inference, an adversary may construct or employ an autonomous agent. 
The design intuition is inspired by how humans reason about image geolocation~\cite{WHMGT18}. 
When given an image, a person might first identify distinctive geographic cues such as landmarks, architectural features, or visible text. 
They may then rely on their own geographic knowledge or perform external web searches for comparison, before combining the evidence to make a final judgment. 

Building on this design intuition, our autonomous agent follows a procedure with four stages when analyzing an image. 
The complete system design is illustrated in~\autoref{fig:architecture}. 

\subsection{Visual Feature Analysis}
\label{sec:visual_feature_analysis}

Inspired by how humans solve geolocation tasks, our agent adapts its strategy to task difficulty: simple images are handled with fewer tools, while complex ones require richer analysis. 
The challenge lies in defining what makes an image easy or difficult to geolocate. 
We propose a weighted heuristic scoring method (\autoref{tab:weighting_scheme}) derived from empirical observations of human geolocation behavior. 
Eight visual cues are identified as key contributors: landmarks, text, architectural style, geographic features, image quality, contextual hints, scene type, and a bonus for multiple cues. 
Landmarks receive the highest weight for their unique discriminative power~\cite{ZZSABBBCN09,WKP16}, followed by textual cues such as place names or local language. 
Architectural and geographic features provide regional but less unique signals, while contextual elements (e.g., vehicles, clothing) and scene type offer auxiliary information. 
Image quality also affects inference reliability, as higher clarity aids fine grained cue extraction. 

An image starts with a base score of 50, adjusted by weighted factors to 100. 
The final score classifies images into five difficulty levels: \textit{Easy (81 to 100)}, \textit{Moderate (61 to 80)}, \textit{Difficult (41 to 60)}, \textit{Very Difficult (21 to 40)}, and \textit{Extremely Difficult (1 to 20)}. 
We use these difficulty levels to guide the agent’s strategy selection. 
Representative examples across difficulty levels are shown in Appendix~\autoref{fig:difficulty}, where higher scores correspond to more distinctive cues, and lower scores to sparse or ambiguous information. 

\subsection{Strategy Execution}
\label{sec:strategy_execution}

After assigning a difficulty score, the agent enters the strategy execution phase, where it decides how to leverage its available tools. 
Instead of following a fixed strategy, the agent adaptively combines the following customized tools to optimize geolocation inference:
(1) LVLMs based analysis, which directly prompts the LVLMs for single step reasoning;
(2) Experience augmented prompting, which applies optimized prompts derived from reasoning patterns observed in past successful cases;
(3) Geographic feature segmentation, which isolates location relevant elements (e.g., landmarks, signage) to enable more precise cue extraction;
(4) Reverse image search, which retrieves visually similar images from the web, filters them via CLIP based similarity, and extracts geographic information from associated metadata and contextual text. 

\mypara{LVLMs Based Analysis}
This module prompts the LVLMs directly with ``Where was the photo taken?'' to infer the specific geolocation.
It is very effective when landmarks or other distinctive features are obvious. 
However, when the image is generic or lacks clear cues, the model is prone to return an ``unknown'' prediction. 

\begin{table*}[!t]
\centering
\caption{Heuristic weighting scheme for visual difficulty assessment.}
\label{tab:weighting_scheme}
\begin{tabular}{l l}
\toprule
\textbf{Factor} & \textbf{Weighting Rule} \\
\midrule
Landmarks & +30 if present \\
Text Visibility & +20 (abundant), +10 (some), +5 (minimal), 0 (none) \\
Architecture & +15 if distinctive \\
Geographic Features & +15 if unique \\
Image Quality & +10 (excellent), +5 (good), 0 (fair), -15 (poor) \\
Contextual Clues & +10 (many), +5 (some), 0 (few), -10 (none) \\
Scene Type & +5 (urban), -5 (rural), -10 (indoor) \\
Bonus for Multiple Cues & +10 if $\geq 3$ indicators, +5 if $\geq 2$ \\
\bottomrule
\end{tabular}
\end{table*}

\mypara{Experience Augemented Prompting}
Naively prompting an LVLM often leads to inconsistent reasoning, as the model may attend to irrelevant details and overlook key geographic cues such as signs or architectural styles. 
To address this, we optimize prompts that preserve general reasoning structure while emphasizing image aware elements, that is, high value cues extracted from the image. 
Semantically aligning prompts with an image’s geographic features helps the LVLM focus on discriminative signals and improve geolocation accuracy. 
GeoCLIP~\cite{CNS23}, which aligns images with GPS based textual descriptions, provides a practical metric for this alignment: higher GeoCLIP similarity indicates that a prompt captures the most distinctive cues. 

We predefine five core categories—architectural, infrastructure, environmental, urban planning, and signage—covering key discriminative elements for geolocation tasks (\autoref{tab:geo_elements} in Appendix). 
For each labeled image, overlapping patches are encoded using GeoCLIP’s image encoder, and candidate elements are encoded with its text encoder. 
Cosine similarity identifies top ranked elements, which are used by the LVLM to generate refined prompts integrating the most relevant cues. 
This process is repeated three times, and the prompt with the highest image to prompt similarity that exceeds the ground truth prompt is selected. 
Optimized prompts are stored in memory and later retrieved for visually similar images, enabling experience augmented reasoning that improves accuracy on challenging cases. 

\mypara{Geographic Feature Segmentation}
This module enhances geolocation accuracy by isolating location relevant cues while reducing visual distractions. 
Direct LVLMs prompting often misses key features, and conventional detectors (e.g., YOLO~\cite{yolov5}) are suboptimal because they rely on fixed object classes and may include irrelevant entities. 
To overcome this, we employ an LVLM based segmentation method that identifies geographically informative elements, predicts bounding boxes, and generates Python code for cropping. 
To preserve context, crops are kept moderately loose. 
Because the boxes produced by the LVLM may be imperfect, the system evaluates them on completeness, centrality, context coverage, and boundary validity, refining boxes iteratively until quality criteria are satisfied. 
This process yields segments that accurately capture geographic cues and improve subsequent retrieval and inference. 

\mypara{Vision Based Reverse Image Search}
Existing reverse search approaches in LVLM agents (e.g., GPT-4o~\cite{GPT4o}, o3~\cite{o3}) convert extracted geographic cues into textual queries for web search, which removes detailed visual information and often results in unreliable matches. 
Our agent instead adopts a more human like strategy: it directly submits the input image to external search engines (e.g., Google, Yandex) to retrieve visually similar candidates, thus preserving full image level information. 
Retrieved results are filtered with GeoCLIP to retain only those exceeding a similarity threshold, ensuring strong visual consistency. 
In practice, modules can be combined for higher robustness. 
For complex images, the agent may first segment geographic regions and then apply reverse image search to these segments, while simultaneously retrieving optimized prompts from memory for similar cases. 
This combined strategy maximizes complementary evidence, as different modules capture distinct geographic cues.
Finally, the enriched report is fed into the LVLM to generate the final structured prediction. 

\subsection{Results Synthesis}
\label{sec:results_synthesis}

After executing its selected strategies, \geoagent enters the result synthesis stage to produce a unified geolocation prediction. 
The output includes hierarchical location information (e.g., country, region, city) and a consolidated explanation summarizing supporting evidence. 

Because different modules generate different forms of output, synthesis adapts accordingly. 
LVLM based analysis and experience augmented prompting yield direct location predictions, requiring only formatting. 
In contrast, reverse image search returns an analysis report, not an immediate prediction. 
For these cases, the agent opens webpages linked to retained images, extracts geographic clues (e.g., captions and metadata), and appends them to the report. 
If multiple sources point to the same location, the agent selects that result; otherwise, a rule based process resolves conflicts by prioritizing explicit place names, number of independent supporting signals, and consistency with visible cues in the input image. 

The report also includes a comparison between the input image and retrieved candidates. 
The LVLM examines each pair to determine whether the scenes match and, if not, identify distinctive geographic elements from the original image. 
Finally, the enriched report is fed into the LVLM to generate the final structured prediction. 

\subsection{Iterative Refinement}
After executing its primary strategy, the agent evaluates its own output without ground truth. 
It checks whether the prediction contains complete geographic information (for example, country, region, and city) and whether the explanation is coherent and supported by evidence.If these conditions are met, the result is accepted as the final output. 

When deficiencies are detected, such as missing predictions, insufficient detail, or weak reasoning, the system invokes fallback strategies through a central planner. 
The planner adaptively selects the next step based on the diagnosed issue. 
For example, if a direct LVLM based analysis yields an overly general result, the planner may trigger experience augmented prompting to leverage prior cases or perform reverse image search to gather external evidence. 
If these also fail, geographic feature segmentation is applied to extract additional clues from focused regions. 

The evaluate then fallback cycle continues iteratively under practical constraints such as time budget. 
It terminates once the agent produces a sufficiently complete answer or the allocated resources are exhausted. 

\section{Experiments}
\label{sec:experiments}

\subsection{Datasets and Metrics}
\label{sec:datasets_and_metrics}

\mypara{Datasets}
We use the MP16-pro dataset~\cite{JLLZWDHWWY24}, which contains over 4 million high quality images with complete geographic annotations. 
In our evaluation, we randomly select 3,000 images to construct the experience augmented module and 1,000 images as the test set. 
We choose this dataset because it is primarily collected from the Flickr platform~\cite{flickr}, where images are captured in everyday contexts and provide a diverse range of visual clues. 
We also test on the DoxBench~\cite{LZLLZXX25} dataset to avoid data contamination. 
This dataset was collected in April 2025 with 500 images across six California cities. 
We defer the discussion of data contamination to~\autoref{section:discussion}. 

\begin{figure*}[!t]
  \centering
  \includegraphics[width=2\columnwidth]{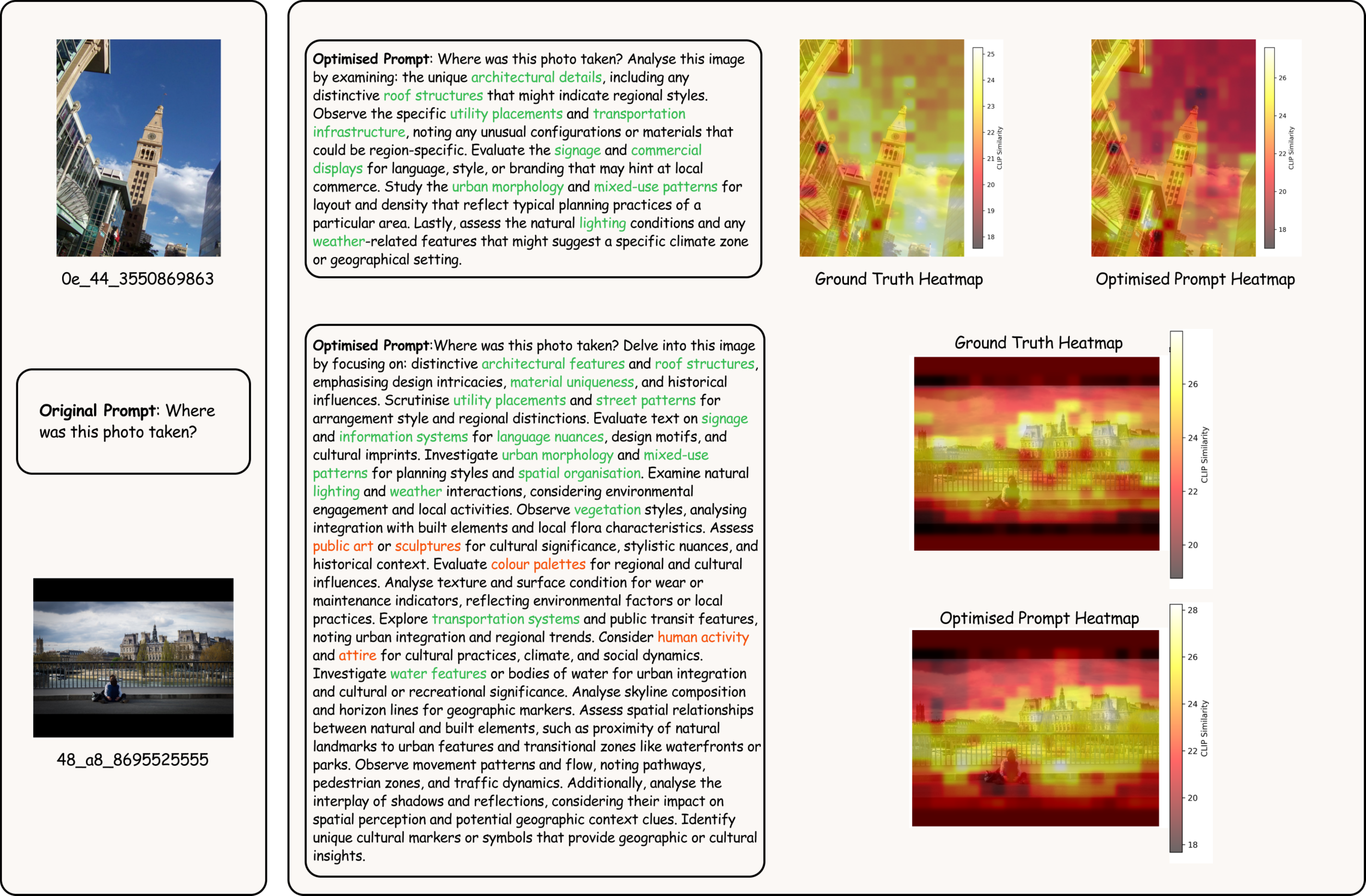}
  \caption{Prompt optimization and GeoCLIP similarity heatmaps. The left column shows the initial prompts, the middle column presents the optimized prompts, and the right column compares the heatmaps generated from the ground-truth prompts and the optimized prompts. 
  The image IDs from the MP16 dataset are displayed at the bottom. In the optimized prompts, elements highlighted in \textcolor{green}{green} correspond to the five predefined categories in Appendix~\autoref{tab:geo_elements} (architectural, infrastructure, environmental, urban planning, signage), while elements in \textcolor{red}{red} denote additional geographic cues introduced by the LLM during the optimization process.}
  \label{fig:heatmap}
\end{figure*}

\mypara{Models}
We evaluate multiple models, including GPT-4o~\cite{GPT4o}, OpenAI o3~\cite{o3}, Gemini 2.5 Pro~\cite{gemini_pro}, and Gemini 2.5 Flash~\cite{gemini_flash}. 
All models are tested under the same evaluation protocol to ensure a fair comparison. 
Note that due to the space limitation, we only include GPT-4o and OpenAI o3 in the main evaluation.
The results of the Gemini series can be seen in Appendix. 

\mypara{Evaluation Method}
In the evaluation stage, determining the correctness of predictions is crucial. 
A straightforward approach is to compare predictions with the ground truth through keyword matching, but this method may introduce statistical errors. 
For example, the predicted result may simultaneously include New York and New York City, or use place names in other languages that differ from the English annotations in the dataset. 
To address these issues, we employ an LVLM as a judge to determine whether the predicted location names are consistent with the ground truth. 
We evaluate the prediction quality along two dimensions:

\begin{itemize}
    \item \textit{Prediction accuracy:} whether the predicted result matches the ground truth at different geographic levels (country, region, city);
    \item
    \textit{Unknown rate:} the proportion of cases where the LVLM outputs ``unknown'' or a similar uncertain answer when making predictions. 
\end{itemize}
It is important to note that we also treat outputs of ``unknown’’ as predictions, and classify them as incorrect cases. 

\subsection{Experience Augmented Module Construction}
\label{sec:experience_augmented_module_construction}

In Section 3.3, we introduced the experience augmented module, which leverages GeoCLIP similarity to enrich prompts with geographic elements and guide the agent’s reasoning toward meaningful cues. 
\autoref{fig:heatmap} shows examples of prompts before and after optimization, along with their GeoCLIP similarity heatmaps. 
Compared with the original versions, the optimized prompts include a broader and more relevant set of geographic elements, encouraging reasoning over diverse visual cues. 
The optimization is not limited to the predefined categories; when other relevant features appear, the LVLM can flexibly incorporate them, extending coverage beyond the core elements. 

To further illustrate this effect, we compare optimized prompts with ground truth prompts derived from true labels (e.g., “This image was taken in Cambridge, Massachusetts, United States”). 
Heatmaps from ground truth prompts exhibit scattered activations, often on uninformative regions such as the sky or background. 
In contrast, optimized prompts yield focused, coherent activations concentrated on discriminative geographic elements—building facades and roofs (Architectural), streets and infrastructure (Infrastructure), and signage or displays (Signage). 
These results demonstrate that optimized prompts improve semantic alignment between text and image, producing more consistent similarity distributions and attention concentrated on regions truly relevant to geolocation. 

\begin{figure*}[!t]
  \centering
 \includegraphics[width=2\columnwidth]{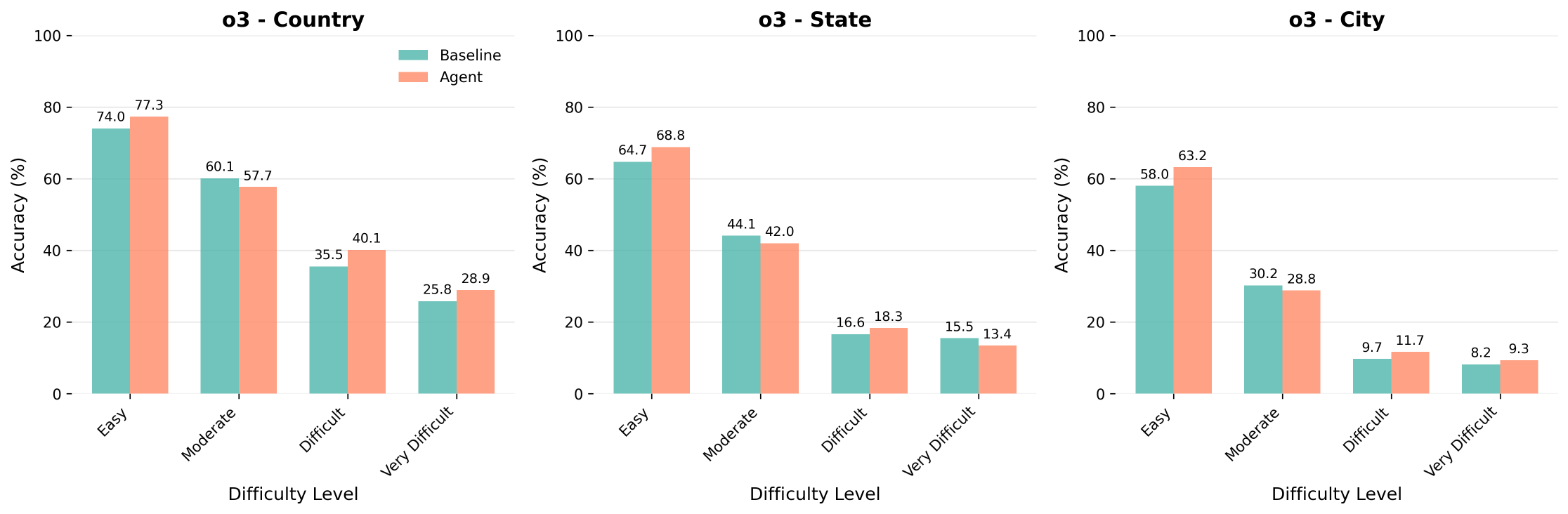}
  \caption{Accuracy comparison of the baseline LVLM and \geoagent across difficulty levels.}
  \label{fig:o3_accuracy_comparison}
\end{figure*}

\subsection{Main Evaluation}
\label{sec:main_evaluation}

First, we aim to assess whether the proposed agent poses greater privacy risks for shared images compared to a baseline LVLM. 
As shown in~\autoref{fig:o3_accuracy_comparison}, \geoagent consistently achieves higher accuracy than the baseline LVLM o3, particularly at higher difficulty levels. 
For instance, at the country level, accuracy in the ``Difficult'' setting increases from 35.5\% to 40.1\%, and in the ``Very Difficult'' setting from 25.8\% to 28.9\%. 
Similar gains appear at the city level, where performance improves from 9.7\% to 11.7\% on ``Difficult'' images and from 8.2\% to 9.3\% on ``Very Difficult'' images. 

In addition, \autoref{tab:o3_unknown} shows that \geoagent significantly reduces the unknown rate for o3 in challenging cases. 
At the ``Difficult'' level, the rate drops from 45.8\% to 22.6\%, and at the ``Very Difficult'' level from 55.7\% to 28.9\%. 
These results demonstrate that the agent can provide more confident predictions when the task becomes challenging. 
Although the proposed agents are not always better than LVLMs on simple tasks, their use of multiple strategies becomes valuable in more difficult cases. 
Meanwhile, this stronger reasoning ability also amplifies privacy risks. 
Unlike baseline LVLMs, the propesed agent combines multiple strategies with external information retrieval, systematically extracting hidden geographic cues. 
These additional signals make it easier for adversaries to infer users' true locations, thereby increasing the threat to user privacy. 

\begin{table}[!t]
\centering
\caption{Unknown rate (\%) of o3 under \geoagent and the baseline across difficulty levels.}
\label{tab:o3_unknown}
\begin{tabular}{l | c | c}
\toprule
& Baseline & \geoagent\\
\midrule
Easy (269)              & 21.9 & 18.6\\
Moderate (281)          & 29.2 & 26.3\\
Difficult (349)         & 45.8 & 22.6\\
Very Difficult (97)     & 55.7 & 28.9\\
Extremely Difficult (4) & 75.0 & 50.0\\
\bottomrule
\end{tabular}
\end{table}

\begin{table*}[!t]
\centering
\caption{Country level accuracy across difficulty levels for o3 (difference relative to baseline).}
\label{tab:o3-country}
\scalebox{0.75}{
\begin{tabular}{l | c | c | c | c | c | c | c}
\toprule
Difficulty (Images) & Baseline & EAP ($\Delta$) & Reverse Search ($\Delta$) & EAP+RS ($\Delta$) & Baseline+Seg+RS ($\Delta$) & EAP+Seg+RS ($\Delta$) & \geoagent ($\Delta$) \\
\midrule
Easy (269)              & 74.0\% & 78.4\% (+4.4) & 76.6\% (+2.6) & 77.0\% (+3.0) & 74.7\% (+0.7) & 75.1\% (+1.1) & 77.3\% (+3.3)\\
Moderate (281)          & 60.1\% & 61.6\% (+1.5) & 50.9\% (-9.2) & 57.3\% (-2.8) & 50.5\% (-9.6) & 51.6\% (-8.5) & 57.7\% (-2.4)\\
Difficult (349)         & 35.5\% & 35.2\% (-0.3) & 37.8\% (+2.3) & 40.4\% (+4.9) & 32.1\% (-3.4) & 33.5\% (-2.0) & 40.1\% (+4.6)\\
Very Difficult (97)     & 25.8\% & 15.5\% (-10.3) & 26.8\% (+1.0) & 33.0\% (+7.2) & 15.5\% (-10.3) & 19.6\% (-6.2) & 28.9\% (+3.1)\\
Extremely Difficult (4) & 0.0\%  & 0.0\% (+0.0)  & 0.0\% (+0.0)  & 0.0\% (+0.0)  & 0.0\% (+0.0)   & 0.0\% (+0.0)   & 0.0\% (+0.0)\\
\bottomrule
\end{tabular}
}
\end{table*}

\begin{table*}[!t]
\centering
\caption{State/Region level accuracy across difficulty levels for o3 (difference relative to baseline).}
\scalebox{0.75}{
\begin{tabular}{l | c | c | c | c | c | c | c}
\toprule
Difficulty (Images) & Baseline & EAP ($\Delta$) & Reverse Search ($\Delta$) & EAP+RS ($\Delta$) & Baseline+Seg+RS ($\Delta$) & EAP+Seg+RS ($\Delta$) & \geoagent ($\Delta$)\\
\midrule
Easy (269) & 64.7\% & 67.3\% (+2.6) & 69.5\% (+4.8) & 71.0\% (+6.3) & 65.1\% (+0.4) & 65.1\% (+0.4) & 68.8\% (+4.1)\\
Moderate (281) & 44.1\% & 43.4\% (-0.7) & 38.1\% (-6.0) & 43.4\% (-0.7) & 35.9\% (-8.2) & 35.9\% (-8.2) & 42.0\% (-2.1)\\
Difficult (349) & 16.6\% & 15.8\% (-0.8) & 17.8\% (+1.2) & 18.9\% (+2.3) & 12.9\% (-3.7) & 16.6\% (+0.0) & 18.3\% (+1.7)\\
Very Difficult (97) & 15.5\% & 7.2\% (-8.3) & 13.4\% (-2.1) & 15.5\% (+0.0) & 7.2\% (-8.3) & 9.3\% (-6.2) & 13.4\% (-2.1)\\
Extremely Difficult (4) & 0.0\% & 0.0\% (+0.0) & 0.0\% (+0.0) & 0.0\% (+0.0) & 0.0\% (+0.0) & 0.0\% (+0.0) & 0.0\% (+0.0)\\
\bottomrule
\end{tabular}
}
\label{tab:o3-state}
\end{table*}

\begin{table*}[!t]
\centering
\caption{City level accuracy across difficulty levels for o3 (difference relative to baseline).}
\label{tab:o3-city}
\scalebox{0.75}{
\begin{tabular}{l | c | c | c | c | c | c | c}
\toprule
Difficulty (Images) & Baseline & EAP ($\Delta$) & Reverse Search ($\Delta$) & EAP+RS ($\Delta$) & Baseline+Seg+RS ($\Delta$) & EAP+Seg+RS ($\Delta$) & \geoagent ($\Delta$)\\
\midrule
Easy (269)              & 58.0\% & 60.6\% (+2.6) & 63.6\% (+5.6) & 63.2\% (+5.2) & 61.3\% (+3.3) & 60.6\% (+2.6) & 63.2\% (+5.2)\\
Moderate (281)          & 30.2\% & 30.2\% (+0.0) & 28.8\% (-1.4) & 30.2\% (+0.0) & 23.5\% (-6.7) & 24.6\% (-5.6) & 28.8\% (-1.4)\\
Difficult (349)         & 9.7\%  & 9.5\% (-0.2)  & 12.3\% (+2.6) & 11.7\% (+2.0) & 8.9\% (-0.8)  & 9.7\% (+0.0)  & 11.7\% (+2.0)\\
Very Difficult (97)     & 8.2\%  & 3.1\% (-5.1)  & 9.3\% (+1.1)  & 8.2\% (+0.0)  & 7.2\% (-1.0)  & 6.2\% (-2.0)  & 9.3\% (+1.1)\\
Extremely Difficult (4) & 0.0\%  & 0.0\% (+0.0)  & 0.0\% (+0.0)  & 0.0\% (+0.0)  & 0.0\% (+0.0)  & 0.0\% (+0.0)  & 0.0\% (+0.0)\\
\bottomrule
\end{tabular}
}
\end{table*}

\subsection{Ablation Study}
\label{sec:ablation_study}

We performed an ablation study to evaluate the contribution of each module and their privacy implications. 
\autoref{tab:o3-country} to~\autoref{tab:o3-city} show the accuracy of GPT o3 at the country, state, and city level geolocation in different difficulty settings. 

The Experience Augmented Prompting (EAP) module shows mixed effects. 
It slightly improves performance in some cases (for example, at the country level under moderate difficulty: from 60.1\% to 61.6\%) but may reduce accuracy when redundant contextual cues add noise (for example, at the country level under very difficult conditions: from 25.8\% to 15.5\%). 
The Reverse Search (RS) module provides the most consistent gains in challenging tasks. 
For example, at the country level under difficult conditions, performance improves from 35.5\% to 37.8\% and and at the city level under difficult conditions from 9.7\% to 12.3\%. 
These results show that external visual references help the model infer location when intrinsic clues are limited.
However, RS can be unstable in moderate cases (at the state level under moderate difficulty, from 44.1\% to 38.1\%), where visually similar but irrelevant matches may mislead reasoning. 
The Segmentation (Seg) module complements RS in difficult settings by isolating geographic subregions and filtering background noise, which improves retrieval reliability. 
For example, at the country level under difficult conditions, accuracy increases from 35.5\% to 40.4\% when segmentation is combined with RS. 

\begin{table*}[!t]
\centering
\caption{Effectiveness of four different defense mechanisms under o3 model.}
\label{tab:4o-o3-all}
\begin{tabular}{l | c | c | c | c}
\toprule
\textbf{Baseline (o3)} & Country & State & City & Unknown\\
\midrule
original       & 50.0\% & 34.0\% & 27.0\% & 33.0\%\\
Watermark      & 6.0\%  & 5.0\%  & 4.0\%  & 94.0\%\\
VPI            & 39.0\% & 31.0\% & 27.0\% & 15.0\%\\
Trigger-based        & 49.0\% & 33.0\% & 29.0\% & 14.0\%\\
EXIF           & 52.0\% & 38.0\% & 27.0\% & 30.0\%\\
\midrule
\textbf{\geoagent (o3)} & Country & State & City & Unknown\\
\midrule
original       & 51.0\% & 34.0\% & 30.0\% & 21.0\%\\
Watermark      & 10.0\% & 6.0\%  & 5.0\%  & 84.0\%\\
VPI            & 41.0\% & 32.0\% & 27.0\% & 17.0\%\\
Trigger-based        & 53.0\% & 37.0\% & 29.0\% & 18.0\%\\
EXIF           & 51.0\% & 34.0\% & 32.0\% & 23.0\%\\
\bottomrule
\end{tabular}
\end{table*}

The agent with autonomous decision making achieves the most consistent improvements. 
For example, at the country level under difficult conditions, the accuracy increases from 35.5\% to 40.1\%, and at the city level under difficult conditions from 9.7\% to 11.7\%. 
These results confirm that single strategies are not sufficient in complex settings, and an adaptive combination of multiple modules yields stronger performance. 
However, this adaptability also makes location-related information easier to capture, amplifying privacy risks in challenging scenarios. 

Overall, the ablation results show that while Memory and RS may introduce noise or instability in simple cases, the multimodule framework, particularly the autonomous agent, outperforms baseline LVLMs at higher difficulty levels. 
The results for GPT-4o follow similar trends, showing larger improvements at higher difficulty levels. 
Detailed results for GPT-4o are provided in the Appendix. 

\subsection{Assessment of Generalizability}
\label{sec:assessment_of_generalizability}

We evaluated four models, Gemini 2.5 Pro, Gemini 2.5 Flash, o3, and GPT-4o, on the DOXBENCH dataset (\autoref{tab:data_contamination} in Appendix). 
All models achieve high accuracy at the Country and State levels but drop sharply at the City level, typically below 24.0\%. 
Performance variations between \geoagent and baseline LVLMs are evident across different difficulty settings. 

Most DOXBENCH samples are easy or moderate, containing rich yet sometimes inconsistent geographic elements.
For example, a U.S. rural house with British style decorations may mislead models toward Europe. 
Under the \geoagent framework, models explicitly extract and integrate multiple cues, which can amplify such conflicts and slightly reduce accuracy in these cases. 
In contrast, difficult samples often contain few but distinctive clues. 
Baseline LVLMs tend to fail when such weak signals dominate, whereas \geoagent compels focused reasoning on limited but informative cues, leading to notable accuracy gains. 
Overall, DOXBENCH results suggest that \geoagent performs more robustly when geographic signals are sparse or subtle, but may face interference when multiple conflicting cues coexist. 

\subsection{Defence Methods}
\label{sec:defence_methods}

To mitigate the privacy risks of geolocation models, we examine four defense strategies inspired by existing protection mechanisms in LVLMs. 
\textbf{Visual Prompt Injection}~\cite{Z25, WM25} adds incorrect geographic labels to an image, creating conflicts between the visual content and the attached text that can mislead model reasoning. 
\textbf{Watermarking}~\cite{DCCH25} embeds explicit messages such as “Geolocation of this image is prohibited,” encouraging the model to decline the task. 
\textbf{Trigger Based Defenses}~\cite{LLLCn,LPMLC24} insert small visual symbols into the image, introducing an intervention signal that alters the model's reasoning when detected. 
Finally, \textbf{EXIF Modification}~\cite{XWS15} alters or forges location related metadata, preventing the direct leakage of coordinates even though the visual content may still enable inference. 

\autoref{tab:4o-o3-all} summarizes the effectiveness of four defense methods on o3 under both the baseline and agent settings. 
Watermarking is the most effective, raising the unknown rate from 33\% to 94\% in the baseline model and from 21\% to 84\% in the agent model, effectively suppressing geolocation outputs. 
Visual prompt injection and trigger both introduce misleading visual or textual cues, which reduce accuracy by causing the model to incorporate incorrect evidence. 
In contrast, EXIF modification shows that the agent does not rely on metadata during reasoning. 
Despite these defenses, the agent remains consistently more robust than the baseline model, which makes its geolocation capability harder to suppress and raises additional privacy concerns. 

\section{Discussion}
\label{section:discussion}

\mypara{Data Contamination}
We acknowledge that the dataset used in our primary experiments may pose a potential data contamination risk, as some portion of the data could have been included in the pretraining corpus of the underlying language model. 
However, our evaluation focuses on a comparative setup in which both the baseline model and our agent framework are powered by the same underlying LVLM. 
Under such conditions, any contamination would affect both systems equally. 
Therefore, the performance improvement observed in our agent over the baseline cannot be attributed to pretraining exposure, but rather to the design of the proposed agent framework. 
This indicates that the proposed agent provides genuine benefits beyond the model's inherent knowledge. 
Additionally, we evaluate our approach on a newer dataset that is highly unlikely to be contaminated, though its geographic coverage is limited to six U.S. cities. 
Future work will focus on expanding the geographic diversity and coverage of such datasets, enabling a more comprehensive assessment. 

\mypara{Future Defense}
We examine several existing defense mechanisms, including visual prompt injection, watermarks, triggers, and EXIF modification. 
Our results show that the proposed agent is more robust to these defenses than standard LVLMs, revealing that an adversary could construct an agent to maximize the geolocation leakage. 
Among these defenses, watermarking is the most effective at forcing the model to refuse geolocation inference, but it inevitably introduces visible artifacts into the image. 
Therefore, we call for stronger defenses that do not degrade the visual quality of images. 

\section{Conclusion}
\label{section:conclusion}

We propose \geoagent, an agentbased system for image geolocation that integrates tools to support multistep reasoning. 
The proposed agent simulates humanlike strategies, extracting subtle geographic cues and leveraging external knowledge to improve inference over baseline LVLMs. 
We evaluate it on MP16-Pro and DoxBench, across multiple advanced LVLMs. 
The evaluation shows that \geoagent achieves higher accuracy in challenging scenarios, reduces the ``unknown'' rate by more than half, and increases risks related to location privacy. 
We study four defense methods and find that watermarking is the most effective, while other methods offer limited protection. 
Overall, \geoagent highlights both the power of agentic reasoning for geolocation and the urgent need for robust privacy defenses. 

\bibliographystyle{plain}
\bibliography{normal_generated}

\clearpage
\appendix

\section{Details of System Design}
\label{sec:detail}

\subsection{Geographic Feature Segmentation}
\label{sec:segementation_example}

\begin{figure}[!h]
    \centering
    \begin{subfigure}{0.48\columnwidth}
        \centering
        \includegraphics[width=\columnwidth]{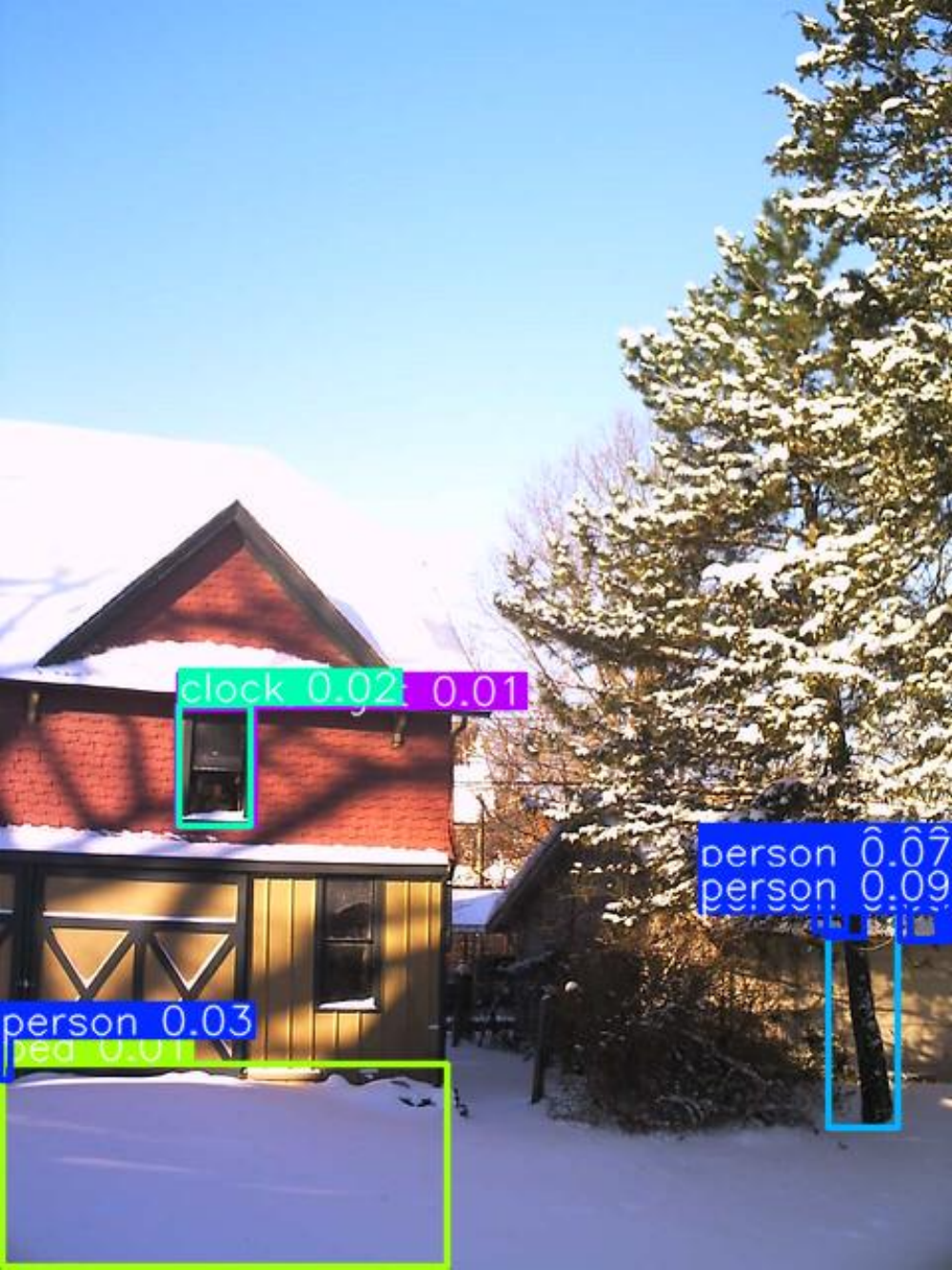}
        \caption{YOLOv5-based segmentation result (conf = 0.01).}
        \label{fig:yolo_segement}
    \end{subfigure}
    \hfill
    \begin{subfigure}{0.48\columnwidth}
        \centering
        \includegraphics[width=\columnwidth]{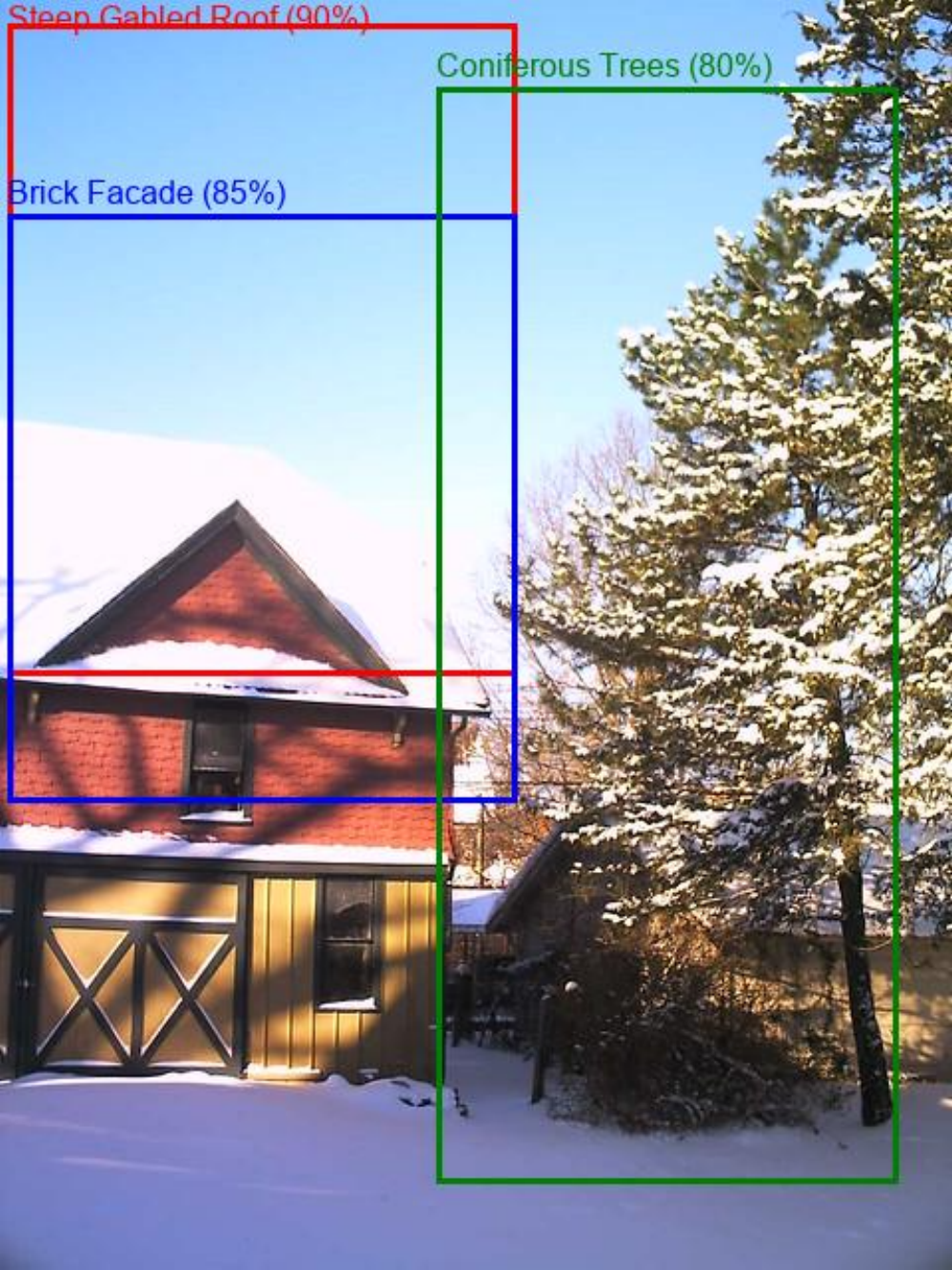}
        \caption{Our geographic feature segmentation module.}
        \label{fig:llm_segeme}
    \end{subfigure}
    \caption{Comparison between our LLM-based geographic feature segmentation (left) and YOLOv5 (right) on an example from the MP16 dataset (image ID: \texttt{1c\_8c\_311262130}).}
    \label{fig:segmentation_example}
\end{figure}

For a direct comparison between YOLOv5 and our geographic feature segmentation module, we present the segmentation results in~\autoref{fig:segmentation_example}, both derived from the image \texttt{1c\_8c\_311262130.jpg} in the MP16-Pro dataset. 

The image on the left shows the segmentation results produced by YOLOv5. 
Our experimental results indicate that under standard confidence thresholds of 0.25, 0.20, and 0.10, the model fails to detect any objects. 
Detections appear only at \text{conf}=0.01, yielding 10 low confidence boxes, all of which are incorrect. 
Further analysis reveals that the image primarily contains natural scene elements such as houses, trees, and snow. 
In contrast, YOLOv5 is trained on the COCO dataset, whose 80 predefined categories do not include buildings, trees, or other outdoor geographic features. 

The right image shows the output of our module: the red box highlights a \textit{steep gabled roof}, the blue box highlights a \textit{brick facade}, and the green box highlights \textit{coniferous trees}. 
The percentages in parentheses indicate the model’s confidence in identifying each feature. 
All detected regions correspond to meaningful geolocation cues. 
Moreover, we intentionally extend the bounding boxes slightly to ensure that each cropped region retains not only the geographic feature itself but also a reasonable amount of surrounding scene context. 
This additional contextual information is essential for improving downstream reverse search on the segmented images. 

\subsection{Vision Based Reverse Image Search}
\label{sec:vision_based_reverse_image_search}

\begin{figure}[!t]
\centering
\includegraphics[width=\columnwidth]{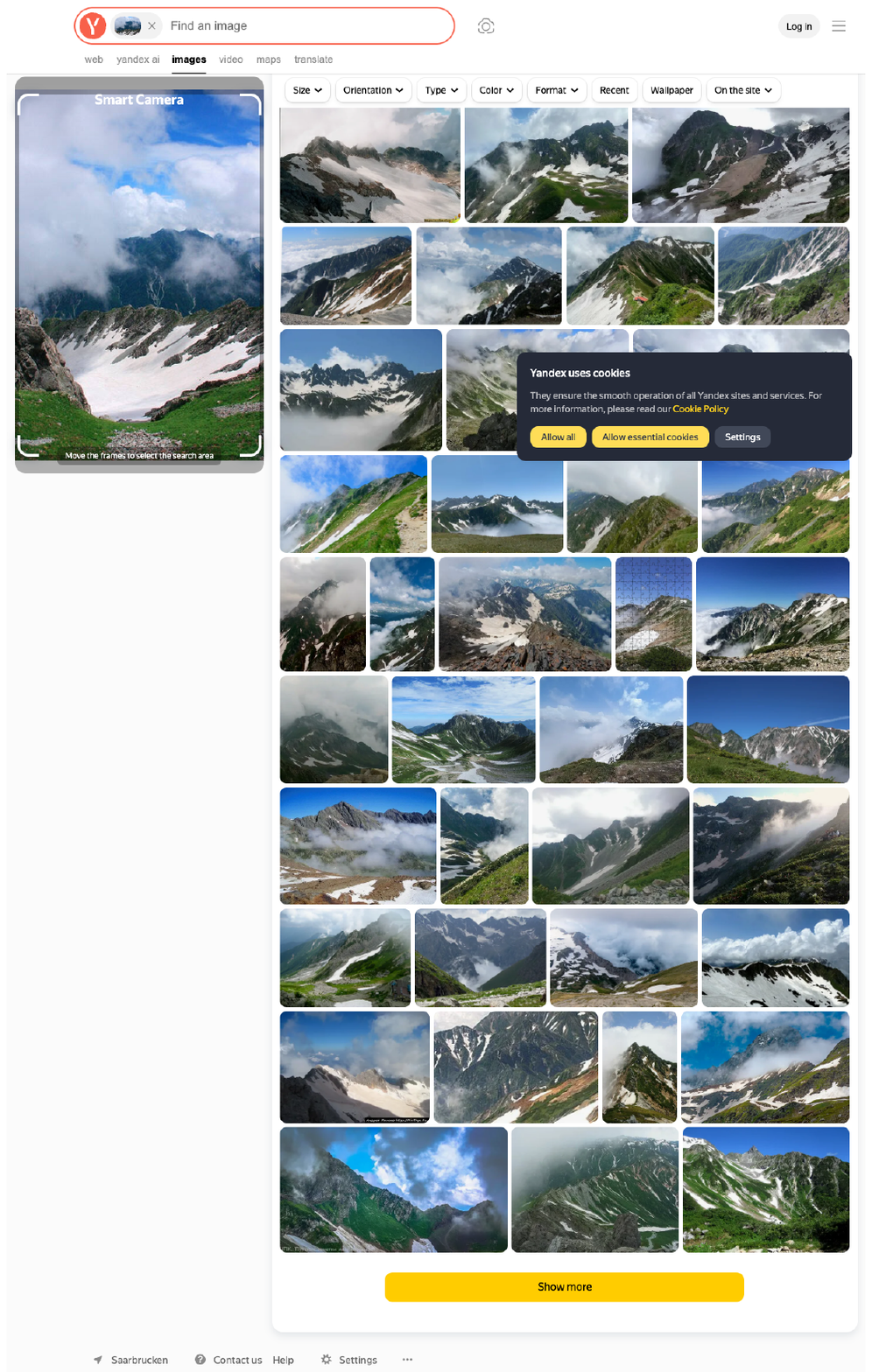}
\caption{Result of simulating human web search operations. (Since automated operations on Google often trigger human verification, Yandex was used for testing instead.)}
\label{fig:search_result}
\end{figure}

\autoref{fig:search_result} shows how we simulate human web searches for reverse image search. 
We package this process into an automated tool. 
Large numbers of automated queries can trigger Google’s security protections. 
Therefore, during testing, we mainly use Yandex’s image search engine. 

\subsection{Prompt Generation}
\label{sec:prompt_detail}

\autoref{tab:geo_elements} provides an overview of the five core categories of geographic elements that guide our cue extraction process. 
These categories cover structural, environmental, and semantic features commonly used by both humans and automated systems to identify geographic context. 
Representative features are listed to illustrate the types of visual cues within each category. 
During cue extraction, these features are encoded as textual candidates and compared with image patches to identify the most relevant matches. 
The selected candidates, together with the original image, are then provided to the LLM to generate a refined prompt that integrates the strongest cues while preserving a generic reasoning structure. 
This process is not strictly limited to the five predefined categories; when distinctive cues fall outside these groups, the LLM may incorporate additional elements. 

\subsection{Difficulty Assessment}
\label{sec:difficulty_detail}

\begin{figure*}[!t]
    \centering
    \includegraphics[width=2\columnwidth]{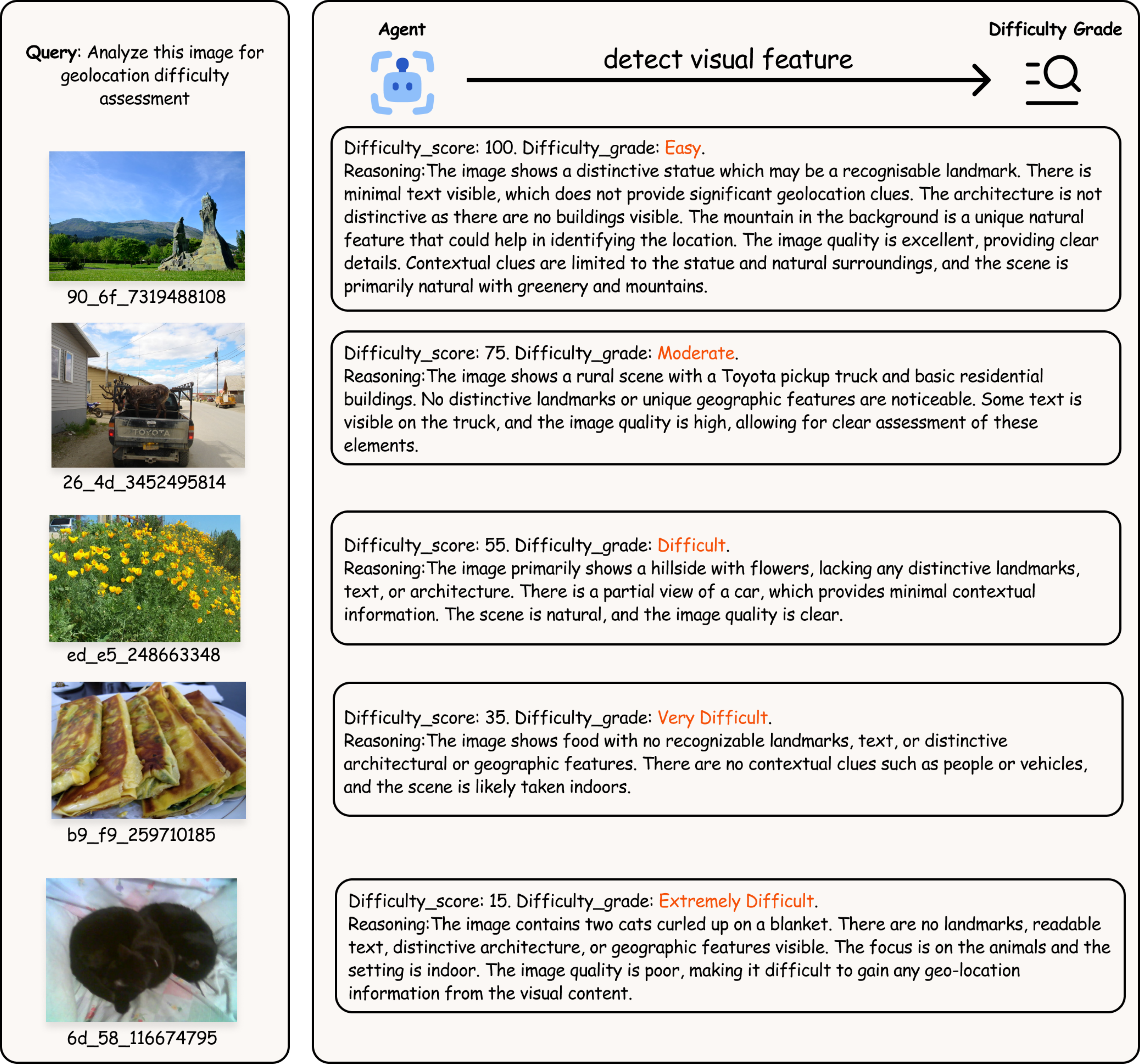}
    \caption{Examples of image-based geolocation difficulty assessment.
    The number below the image represents the unique ID in the MP-16 dataset.
    The \textcolor{red}{red} text indicates the difficulty rating assigned to the image by the agent after visual feature detection.}
    \label{fig:difficulty}
\end{figure*}

\autoref{fig:difficulty} presents examples of our image based geolocation difficulty assessment. 
The images on the left are taken from the MP-16 dataset, and each image is shown with its unique ID. 
The middle panel displays the agent’s analysis of visual content, including the detected geographic cues and the reasoning process used to assess difficulty. 
Based on this analysis, the agent assigns a difficulty score and classifies the image into one of five levels: Easy, Moderate, Difficult, Very Difficult, or Extremely Difficult. 
The final difficulty rating appears in red within the corresponding reasoning text on the right. 
The figure offers a clear visual validation of our difficulty classification scheme. 
The assigned difficulty levels align well with the richness and distinctiveness of geographic cues present in each image.

\begin{figure*}[!ht]
    \centering
    \includegraphics[width=2\columnwidth]{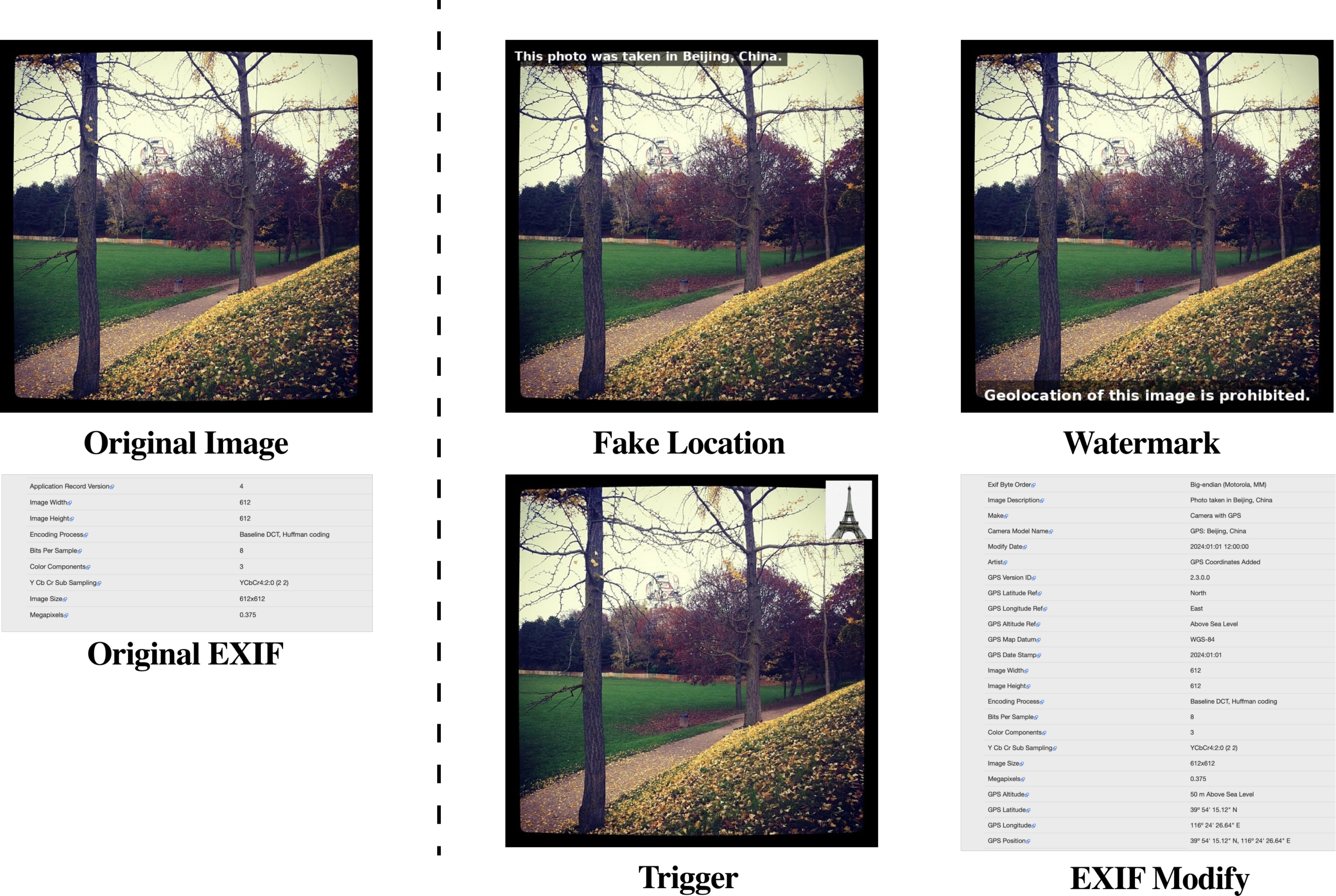}
    \caption{Examples of four defense methods against geolocation privacy attacks.
    The left side shows the original image and its EXIF information (without any location data).
    The right side illustrates four defenses: (1) Fake Location: inserting misleading location descriptions; (2) Watermark: overlaying explicit warnings against geolocation; (3) Trigger: adding distracting visual elements to confuse recognition; and (4) EXIF Modify: altering or forging location-related metadata.}
    \label{fig:defence}
\end{figure*}

\begin{table*}[!t]
\centering
\caption{Five core categories of geographic elements with representative sub-elements.}
\label{tab:geo_elements}
\scalebox{0.75}{%
\begin{tabular}{l | c}
\toprule
Core Element & Example elements\\
\midrule
Architectural & Facades; construction materials (brick, stone, wood); roof styles; window designs; structural systems\\
Infrastructure & Street layouts; pavement materials; transportation systems (rail, tram lines); utility placements (power lines, canals)\\
Environmental & Vegetation types (conifers, palm trees); climate adaptations; natural water bodies (rivers, lakes, coasts); lighting and seasonal variations\\
Urban Planning & Development density; mixed-use patterns; city morphology (grid, radial); land-use arrangements\\
Signage & Traffic signs; billboards; shop signs; typography; symbolic systems (multilingual labels, scripts)\\
\bottomrule
\end{tabular}
}
\end{table*}

\section{Additional Experimental Results}
\label{sec:additional_experimental_results}

\subsection{Ablation Study of GPT-4o}
\label{sec:ablation_study_4o}

\autoref{tab:4o-country} to~\autoref{tab:4o-city} reports the geolocation accuracy of GPT-4o across different difficulty levels. 
The Experience Augmented Prompting (EAP) module shows limited effectiveness. 
For example, at the country task under the easy difficulty, accuracy decreases from 86.6\% to 67.7\%. 
This suggests that additional cues may introduce noise when the image already contains sufficient geographic information. 

The Reverse Search (RS) module is most effective on harder tasks for 4o. 
For example, at the country task under the very difficult difficulty, accuracy increases from 7.2\% to 24.7\%. 
This shows that external visual references help the model infer location when intrinsic cues in the image are sparse. 

The Segmentation (Seg) module supports RS when tasks are judged to be more difficult. 
For example, at the country task under the moderate difficulty, accuracy increases from 50.9\% to 59.1\%. 
This indicates that segmentation helps isolate informative regions and reduces background noise, which improves the reliability of retrieved references.

The Agent with autonomous reasoning provides the most stable improvements for GPT-4o. 
For example, at the country task under the difficult difficulty, accuracy increases from 8.0\% to 26.6\%. 
This shows that the agent can combine multiple signals more effectively than individual modules. 

Overall, the ablation study shows that some modules may add noise in simple cases. 
However, at higher difficulty levels, the combination of modules, especially the Agent with autonomous reasoning, leads to substantial performance gains on GPT-4o. 

\begin{table*}[!t]
\centering
\caption{Country-level accuracy across difficulty levels for 4o (difference relative to baseline).}
\label{tab:4o-country}
\scalebox{0.75}{
\begin{tabular}{l | c | c | c | c | c | c | c}
\toprule
Difficulty (Images) & Baseline & EAP ($\Delta$) & Reverse-Search ($\Delta$) & EAP+RS ($\Delta$) & Baseline+Seg+RS ($\Delta$) & EAP+Seg+RS ($\Delta$) & Geo-Detective ($\Delta$) \\
\midrule
Easy (269)              & 86.6\% & 67.7\% (-18.9) & 84.8\% (-1.8) & 82.5\% (-4.1) & 84.8\% (-1.8) & 83.3\% (-3.3) & 74.7\% (-11.9) \\
Moderate (281)          & 50.9\% & 32.4\% (-18.5) & 55.9\% (+5.0) & 49.1\% (-1.8) & 59.1\% (+8.2) & 49.1\% (-1.8) & 40.9\% (-10.0) \\
Difficult (349)         & 20.1\% & 8.0\% (-12.1)  & 31.2\% (+11.1) & 25.2\% (+5.1) & 30.9\% (+10.8) & 22.9\% (+2.8) & 26.6\% (+6.5) \\
Very Difficult (97)     & 7.2\%  & 0.0\% (-7.2)   & 24.7\% (+17.5) & 19.6\% (+12.4) & 17.5\% (+10.3) & 12.4\% (+5.2) & 15.5\% (+8.3) \\
Extremely Difficult (4) & 0.0\%  & 0.0\% (+0.0)   & 25.0\% (+25.0) & 0.0\% (+0.0)  & 0.0\% (+0.0)  & 0.0\% (+0.0)  & 0.0\% (+0.0) \\
\bottomrule
\end{tabular}
}
\end{table*}

\begin{table*}[!t]
\centering
\caption{State-level accuracy across difficulty levels for 4o (difference relative to baseline).}
\label{tab:4o-state}
\scalebox{0.75}{
\begin{tabular}{l | c | c | c | c | c | c | c}
\toprule
Difficulty (Images) & Baseline & EAP ($\Delta$) & Reverse-Search ($\Delta$) & EAP+RS ($\Delta$) & Baseline+Seg+RS ($\Delta$) & EAP+Seg+RS ($\Delta$) & Geo-Detective ($\Delta$)\\
\midrule
Easy (269)              & 73.2\% & 57.6\% (-15.6) & 70.6\% (-2.6) & 70.3\% (-2.9) & 72.5\% (-0.7) & 67.7\% (-5.5) & 62.8\% (-10.4)\\
Moderate (281)          & 35.2\% & 22.1\% (-13.1) & 34.5\% (-0.7) & 32.0\% (-3.2) & 35.2\% (+0.0) & 33.1\% (-2.1) & 24.6\% (-10.6)\\
Difficult (349)         & 8.9\%  & 5.2\% (-3.7)   & 12.0\% (+3.1) & 10.3\% (+1.4) & 13.8\% (+4.9) & 10.3\% (+1.4) & 10.9\% (+2.0)\\
Very Difficult (97)     & 5.2\%  & 0.0\% (-5.2)   & 9.3\% (+4.1) & 9.3\% (+4.1)  & 7.2\% (+2.0)  & 6.2\% (+1.0) & 8.2\% (+3.0)\\
Extremely Difficult (4) & 0.0\%  & 0.0\% (+0.0)   & 0.0\% (+0.0) & 0.0\% (+0.0)  & 0.0\% (+0.0)  & 0.0\% (+0.0) & 0.0\% (+0.0)\\
\bottomrule
\end{tabular}
}
\end{table*}

\begin{table*}[!t]
\centering
\caption{City-level accuracy across difficulty levels for 4o (difference relative to baseline).}
\label{tab:4o-city}
\scalebox{0.75}{
\begin{tabular}{l | c | c | c | c | c | c | c}
\toprule
Difficulty (Images) & Baseline & EAP ($\Delta$) & Reverse-Search ($\Delta$) & EAP+RS ($\Delta$) & Baseline+Seg+RS ($\Delta$) & EAP+Seg+RS ($\Delta$) & Geo-Detective ($\Delta$)\\
\midrule
Easy (269)              & 62.5\% & 49.4\% (-13.1) & 61.0\% (-1.5) & 60.6\% (-1.9) & 61.3\% (-1.2) & 59.1\% (-3.4) & 56.5\% (-6.0)\\
Moderate (281)          & 22.4\% & 14.2\% (-8.2)  & 21.0\% (-1.4) & 20.6\% (-1.8) & 22.1\% (-0.3) & 21.0\% (-1.4) & 14.6\% (-7.8)\\
Difficult (349)         & 5.4\%  & 3.7\% (-1.7)   & 8.0\% (+2.6)  & 6.3\% (+0.9)  & 7.7\% (+2.3)  & 7.7\% (+2.3)  & 7.2\% (+1.8)\\
Very Difficult (97)     & 4.1\%  & 0.0\% (-4.1)   & 4.1\% (+0.0)  & 4.1\% (+0.0)  & 7.2\% (+3.1)  & 4.1\% (+0.0)  & 5.2\% (+1.1)\\
Extremely Difficult (4) & 0.0\%  & 0.0\% (+0.0)   & 0.0\% (+0.0)  & 0.0\% (+0.0)  & 0.0\% (+0.0)  & 0.0\% (+0.0)  & 0.0\% (+0.0)\\
\bottomrule
\end{tabular}
}
\end{table*}

\subsection{Assessment of Generalizability}
\label{sec:generalizability_results}

To evaluate the generalizability of the agent, we tested four models (Gemini 2.5 Pro, Gemini 2.5 Flash, o3, and GPT-4o) on the DOXBENCH dataset, and the results are shown in Table 11. 

\subsection{Defence Methods}
\label{sec:defence_detail}

\autoref{fig:defence} shows our four types of defense strategies:

\begin{itemize}
    \item Visual Prompt Injection: Add incorrect geographic labels to an image, for example marking a photo taken in Paris as “Beijing, China.” 
    This creates an information conflict between the visible content of the image and its attached text label, which can mislead the model’s reasoning and reduce the reliability of geolocation predictions.
    \item Watermark: Embed explicit messages in the image, such as “Geolocation of this image is prohibited.” This approach is similar to prompt injection, where added input constraints guide the model to refuse geolocation tasks.
    \item Trigger based Defense: Insert specific visual symbols into the image, such as a small landmark icon. 
    This places an intervention signal at the visual level, so that the model changes its reasoning process when the trigger element is detected.
    \item EXIF Modification: Alter or forge the geographic metadata in the image file to prevent direct leakage of geolocation. 
    For instance, replace the true coordinates with random values. 
    This blocks location exposure at the metadata level, though the visual content of the image can still be used for inference.
\end{itemize}

\autoref{tab:4o-all} shows that GPT-4o reacts very differently to each defense. 
Watermark works best. 
In the baseline model, the unknown rate rises from 54\% to 74\%. 
In the agent model, it falls to 17\%, which means the watermark signal becomes weak during iterative reasoning.
VPI and trigger mislead the model. 
In the baseline setting, VPI gives an unknown rate of 25\% and a city accuracy of 22\%. 
Trigger gives 9\% unknown and 28\% city accuracy. 
The agent version is still misled but loses less accuracy.
EXIF modification has almost no effect. 
The baseline city accuracy stays at 26\% and the unknown rate stays at 50\%. 
The agent model is similar.Overall, watermark blocks predictions, VPI and trigger reduce accuracy, and EXIF does not change GPT-4o’s behavior. 

\begin{table*}[!t]
\centering
\caption{Performance comparison between the baseline LVLM and GEO-Detective across four models.}
\label{tab:data_contamination}
\begin{tabular}{l | c c c | c c c}
\toprule
& \multicolumn{3}{c|}{Baseline LVLM} & \multicolumn{3}{c}{GEO-Detective}\\
\midrule
Gemini 2.5 Pro & Country & State & City & Country & State & City\\
\midrule
Easy (117)     & 100.0\% & 89.7\% & 15.4\% & 100.0\% & 89.7\% & 17.1\%\\
Moderate (365) & 99.2\%  & 92.6\% & 16.7\% & 98.4\%  & 91.0\% & 15.3\%\\
Difficult (18) & 94.4\%  & 55.6\% & 5.6\%  & 100.0\% & 55.6\% & 5.6\%\\
\midrule
Gemini 2.5 Flash & Country & State & City & Country & State & City\\
\midrule
Easy (117)     & 100.0\% & 89.7\% & 13.7\% & 100.0\% & 91.5\% & 15.4\%\\
Moderate (365) & 98.9\%  & 89.3\% & 18.4\% & 97.3\%  & 86.6\% & 17.8\%\\
Difficult (18) & 83.3\%  & 50.0\% & 5.6\%  & 88.9\%  & 50.0\% & 5.6\%\\
\midrule
o3 & Country & State & City & Country & State & City\\
\midrule
Easy (117)     & 72.6\% & 64.1\% & 17.1\% & 72.6\% & 63.2\% & 17.1\%\\
Moderate (365) & 85.2\% & 79.2\% & 24.9\% & 79.2\% & 73.7\% & 22.2\%\\
Difficult (18) & 61.1\% & 38.9\% & 11.1\% & 72.2\% & 44.4\% & 11.1\%\\
\midrule
GPT-4o & Country & State & City & Country & State & City\\
\midrule
Easy (117)     & 99.1\% & 88.9\% & 29.1\% & 90.6\% & 81.2\% & 23.9\%\\
Moderate (365) & 88.5\% & 83.3\% & 30.7\% & 78.1\% & 71.8\% & 23.6\%\\
Difficult (18) & 83.3\% & 44.4\% & 5.6\%  & 100.0\% & 55.6\% & 11.1\%\\
\bottomrule
\end{tabular}
\end{table*}

\begin{table*}[!t]
\centering
\caption{Evaluation of defense effectiveness across four methods (original, watermark, visual prompt injection (VPI), exif, and trigger) under GPT-4o models, considering both the baseline and agent settings.}
\label{tab:4o-all}
\begin{tabular}{l | c | c | c | c}
\toprule
Baseline (GPT-4o) & Country & State & City & Unknown\\
\midrule
original       & 40.0\% & 32.0\% & 26.0\% & 54.0\%\\
Watermark      & 26.0\% & 22.0\% & 19.0\% & 74.0\%\\
VPI            & 32.0\% & 25.0\% & 22.0\% & 25.0\%\\
Trigger-based  & 43.0\% & 33.0\% & 28.0\% & 9.0\%\\
EXIF           & 41.0\% & 33.0\% & 26.0\% & 50.0\%\\
\midrule
\geoagent (GPT-4o) & Country & State & City & Unknown\\
\midrule
original       & 42.0\% & 32.0\% & 27.0\% & 10.0\%\\
Watermark      & 38.0\% & 25.0\% & 21.0\% & 17.0\%\\
VPI & 33.0\% & 23.0\% & 22.0\% & 12.0\%\\
Trigger-based  & 41.0\% & 30.0\% & 28.0\% & 16.0\%\\
EXIF           & 46.0\% & 33.0\% & 28.0\% & 10.0\%\\
\bottomrule
\end{tabular}
\end{table*}

\end{document}